\documentclass[intlimits,twoside,a4paper]{article}

\usepackage{amsmath,amssymb}
\usepackage{graphicx}

\usepackage[T2A]{fontenc}
\usepackage[cp1251]{inputenc}
\usepackage[eqsecnum]{cmpj2}

\issue{2013}{16}{4}{43606}
\doinumber{10.5488/CMP.16.43606}

\title[Closed-loop liquid-liquid immiscibility]%
{Closed-loop liquid-liquid immiscibility in mixture of particles with spherically symmetric interaction}
\author[Yu.V. Kalyuzhnyi, T.V. Hvozd]{Yu.V. Kalyuzhnyi, T.V. Hvozd}
\address{Institute for Condensed Matter Physics of the National Academy of Sciences of Ukraine, \\ 1~Svientsitskii St., 79011 Lviv, Ukraine
}

\date{Received August 8, 2013, in final form September 9, 2013}
\authorcopyright{Yu.V. Kalyuzhnyi, T.V. Hvozd, 2013}

\begin{document}

\maketitle

\begin{abstract}

Thermodynamic perturbation theory for central-force (TPT-CF) type of associating potential is used to study
the phase behavior of symmetric binary mixture of associating particles with spherically symmetric
interaction. The model is represented by the binary Yukawa hard-sphere mixture with additional spherically
symmetric square-well associative interaction located inside the hard-core region and valid only between
dissimilar species. To account for the change of the system packing fraction due to association we propose
an extended version of the TPT-CF approach. In addition to the already known four types of the phase diagram
for binary mixtures we were able to identify the fifth type, which is characterized by the absence of
intersection of the $\lambda$-line with the liquid-vapour binodals and by the appearance of the closed-
loop liquid-liquid immiscibility with upper and lower critical solution temperatures.

\keywords thermodynamic perturbation theory, liquid-vapour coexistence, demixing, binary mixture, associating fluids

\pacs 64.70.Ja, 05.70.Jk, 82.70.Dd, 64.10.+h

\end{abstract}

\section{Introduction}

According to the Gibbs phase rule, the binary mixture may have up to four coexisting phases simultaneously.
This fact implies that the phase behavior of the binary fluid could be very rich and complicated. Systematic
study and classification of the peculiarities of the binary systems phase diagrams topologies has been
undertaken more than 40 years ago by Scott and van Konynenburg \cite{Scott1,Scott2}. These studies are
based on the application of the van der Waals equation of state, which in most cases is capable of providing
qualitatively correct description of the phase behavior. Most of the subsequent studies, carried out using
quantitatively more accurate methods of the modern liquid state theory \cite{mcdonald}, have been focused on
the investigation of the phase behavior of symmetric binary mixtures
\cite{wilding,kahl1,kahl2,foffi}. These are the mixtures with identical interaction between particles
of the similar species and different interaction between particles of the dissimilar species.
Phase behavior of such mixtures is defined by the competition between gas-liquid and liquid-liquid coexistence.
Combining Monte-Carlo computer simulation and theoretical mean-field calculations, Wilding et al.
\cite{wilding} identified three types of a phase diagram for square-well hard-sphere
symmetrical binary fluid mixture. Similar three types of a phase diagram were detected in the symmetrical
binary hard-sphere Yukawa mixture using self-consistent Ornstein-Zernike approximation (SCOZA) and Monte-Carlo computer
simulation method \cite{kahl1,kahl2}. At the same time, the phase diagram of the fourth type was also
detected using SCOZA approach \cite{kahl3}.
More recently, the first-order thermodynamic perturbation theory has been used
to study the phase behavior of the binary Yukawa mixture with asymmetry in hard-sphere sizes \cite{foffi}.

In this study we are focused on the investigation of the phase behavior of symmetric Yukawa hard-sphere
binary mixture with additional spherically symmetric square-well associative interaction located inside the
 hard-core region. This additional interaction is valid only between particles of dissimilar
species. The hard-sphere version of the model has been developed and studied by Cummings and Stell
\cite{cummingsstell} and by Kalyuzhnyi et al. \cite{kal_dim}.
Originally, this version of the model was used as a simple hamiltonian model of the chemical
reaction \cite{cummingsstell}. On the other hand, the model of this type can be regarded as
a coarse grained version of the model for
sterically or charge stabilized colloidal dispersions, protein solutions, star-polymer fluids, etc.
\cite{wl10,wl15,wl16}.
Effective interaction between macroparticles of such systems has an attractive potential well at short distances
and a repulsive potential mound at intermediate distances.
The phase diagrams, which include two-phase gas-liquid, liquid-liquid
diagrams and three-phase gas-demixed liquid diagrams have been calculated using thermodynamic perturbation
theory for central force associating potential \cite{tpt1,tpt2,tpt3}.

 Phase behavior of the model, which is
similar to the present one has been studied earlier by Jakson \cite{jackson}. His model is represented
by the symmetrical binary hard-sphere mixture with mean-field type of attractive interaction valid only
between the same species and orientationally dependent associative interaction between particles of unlike
species. Associative interaction appears due to the off-center square-well sites. This model was used as a
generic model for the phase behavior description of the binary mixtures with the possibility of hydrogen
bond formation between unlike species. Combining Wertheim's TPT for associating fluids and mean-field approach,
Jackson was able to show that for a certain choice of the potential model parameters, the system exhibits the
closed loop liquid-liquid immiscibility with the upper and lower critical solution temperatures. It was
concluded that closed loop coexistence appeares due to the presence of the highly anysotropic attraction
between off-center bonding sites.

In the present work we demonstrate that the systems with spherically symmetric interaction may also have closed loop liquid-liquid immiscibility. The paper is organized as follows. In section~2 we describe the model to
be considered and in section~3 we present and discuss details of the TPT-CF
theory, specialized to the model at hand. Our results and discussion are included in section~4 and our
conclusions are collected in section~5.

\section{The model}

We consider symmetric binary Yukawa hard-sphere mixture with additional associative interaction between
dissimilar particles. The total pair potential of the model $U_{ij}(r)$ is represented as a sum of hard-sphere
Yukawa potential $U_{ij}^\textrm{HSY}(r)$ and associative potential $U_\textrm{ass}(r)$, i.e.:

\begin{equation}
\label{U}
U_{ij}(r)=U_{ij}^\textrm{HSY}(r)+(1-\delta_{ij})U_\textrm{ass}(r),
\end{equation}
where the lower indices $i,j$ denote the species of the particles and $\delta_{ij}$ is the Kroneker delta.
In our symmetric binary system, Yukawa interaction between particles of the same species is the same, i.e.,
$U_{11}^\textrm{HSY}(r)=U_{22}^\textrm{HSY}(r)$, and between the particles of dissimilar species it is regulated
by the parameter $\alpha$ ($0<\alpha<1$), i.e. $U_{12}^\textrm{HSY}(r)=\alpha U_{11}^\textrm{HSY}(r)$. We have:
\begin{equation}
\label{UHC1}
  U_{ii}^\textrm{HSY}(r)=\left\{\begin{array}{ll}
                       \infty, & \hbox{$r\leqslant d_{ii}$}\,, \\
                       -\frac{\epsilon_0}{r}\frac{A_{ii}}{z_n}\re^{-z_n(r-d_{ii})} , & \hbox{$r>d_{ii}$}\,,
                      \end{array}
                   \right.
\end{equation}
\begin{equation}
\label{UHC2}
  U_{12}^\textrm{HSY}(r)=U_{21}^\textrm{HSY}(r)=\left\{\begin{array}{ll}
                       \infty, & \hbox{$r<L-\omega/2$}\,,\\
  A_0, & \hbox{$L-\omega/2<r<d_{12}$}\,,\\
  -\frac{\epsilon_0}{r}\frac{A_{12}}{z_n}\re^{-z_n(r-d_{12})}, & \hbox{$r>d_{12}$}\,,
                      \end{array}
                   \right.
\end{equation}
where $A_{11}=A_{22}=dz_n$, $A_{12}=A_{21}=\alpha A_{11}$,
$z_n$ and $\epsilon_0$ are the screening length and the interaction strength of the
Yukawa potential, respectively, $d_{ij}=(d_i+d_j)/2$, $d_i$ is the hard-sphere diameter.
We consider the system
with hard spheres of equal size, i.e., $d_1=d_2=d$.
In (\ref{U})
\begin{equation}
\label{ass}
U_\textrm{ass}(r)=\left\{\begin{array}{ll}
                  0 \,, &   \hbox{$r<L-\omega/2$}\, , \\
                  -\varepsilon_\textrm{ass}-A_0\, , & \hbox{$L-\omega/2<r<L+\omega/2$}\,, \\
                  0 \,,   & \hbox{$r>L+\omega/2$}\,,
                      \end{array}
                   \right.
\end{equation}
where $L$ is the bonding distance, $\omega$ and $\varepsilon_\textrm{ass}$ are the square well potential width and
depth, respectively. In what follows we will consider the hard-sphere Yukawa potential (\ref{UHC1}) and
(\ref{UHC2}) in the limit of $A_0\rightarrow\infty$, and associative potential (\ref{ass})
in the limit of sticky interaction
under the condition that the second virial coefficient remains unchanged.
In this limit, the Mayer function for associative potential $f_\textrm{ass}(r)=\exp{[-\beta U_\textrm{ass}(r)]}-1$ is
substituted by the Dirac delta-function, i.e., $e_{12}^{(\textrm{HSY})}(r)f_\textrm{ass}(r)\rightarrow B\delta(r-L)$, where
$e_{12}^{(\textrm{HSY})}(r)=\exp{[-\beta U_{12}^{(\textrm{HSY})}(r)]}$ and
\begin{equation}
\label{B}
B=L^{-2}\int r^2e_{12}^{(\textrm{HSY})}(r)f_\textrm{ass}(r)\;\rd r.
\end{equation}

The type of the clusters, which will be formed in the system due to association depends on the value of the
bonding length $L$ \cite{tpt1,kal_dim}. For values of L lying in the interval $(0,d/2)$ only dimers can be
formed in the system. When $d/2 < L < \sqrt{3}d$, the formation of the chains is possible.
Further increase in $L$ leads to an increase in the maximum number of the particles of the type 1 (or 2),
which can be simultaneously associated with the particle of the type 2 (or 1), and the formation of the branched
chains will be possible.

The mixture is characterized by the temperature $T$ (or $\beta=(k_\textrm{B}T)^{-1}$, where $k_\textrm{B}$ is the Boltzmann's
constant), the total number-density $\rho$, and the mole (number) fraction $x$ of species $1$ ($x=x_1$); partial number
densities are defined via $\rho_1=x\rho$ and $\rho_2=(1-x)\rho$. We further introduce the reduced
dimensionless quantities, $\rho^*=\rho d^3$, $T^*=k_\textrm{B}T/\epsilon_0$ and
$\epsilon_\textrm{ass}^*=\epsilon_\textrm{ass}/\epsilon_0$.

\section{Theory}

To describe thermodynamic properties of the model at hand we will utilize here thermodynamic perturbation
theory for central force associative potential (TPT-CF) \cite{tpt1,tpt2,tpt3}. According to TPT-CF, Helmholtz
free energy of the system $A$ can be written as a sum of two terms: free energy of the reference system
$A_\textrm{ref}$ and the term describing the contribution to the free energy due to association $A_\textrm{ass}$:

\begin{equation}
\label{FRE total}
 A=A_\textrm{ref}+A_\textrm{ass}=A_\textrm{HSY}+A_\textrm{ass}\,.
\end{equation}
Here, $A_\textrm{ref}=A_\textrm{HSY}$, where $A_\textrm{HSY}$ is the free energy of the hard-sphere Yukawa fluid. To calculate
$A_\textrm{HSY}$, we are using the high temperature approximation. All the rest of thermodynamical quantities
can be obtained using the expression for Helmholtz free energy (\ref{FRE total}) and standard thermodynamical
relations, e.g., differentiating $A$ with respect to the density, we get the expression for the chemical potential:
\begin{equation}
\label{chem total}
\beta\mu_k=\frac{\partial}{\partial\rho_k}\bigg(\frac{\beta A}{V}\bigg),
\end{equation}
and the expression for the pressure $P$ of the system can be calculated invoking the following general
relation:
\begin{eqnarray}
\label{pressure_}
\beta P=\beta \sum_{k}\rho_k\mu_k-\frac{\beta A}{V}\,.
\end{eqnarray}

\subsection{High temperature approximation}

Under the high temperature approximation, the expression for the free energy is:
\begin{equation}
\label{FRE}
\frac{\beta A_\textrm{HSY}}{V}=\frac{\beta A_\textrm{HS}}{V}+2\pi\beta\sum_{i}\sum_{j}\rho_i\rho_j\int_{0}^{\infty} \rd rr^2U_{ij}^\textrm{HSY}(r)g_\textrm{HS}(r),
\end{equation}
where $A_\textrm{HS}$ is the hard-sphere Helmholtz free energy and
$g_\textrm{HS}(r)$ is the hard-sphere radial distribution function. Substituting into (\ref{FRE}) the expression
for the potential (\ref{UHC1}) and (\ref{UHC2}), we have
\begin{equation}
\label{FRE1}
\frac{\beta A_\textrm{HSY}}{V}=\frac{\beta A_\textrm{HS}}{V}-{2\pi\beta\epsilon_0\over z_n}
\widetilde{G}_\textrm{HS}(z_n)\sum_{i}\sum_{j}\rho_i\rho_jA_{ij}\,,
\end{equation}
where $\widetilde{G}_\textrm{HS}(z_n)$ is the Laplace transform of hard-sphere radial distribution function
\begin{equation}
\label{Gtrdef}
\widetilde{G}_\textrm{HS}(z_n)=\re^{z_nd_{ij}}\int_{0}^{\infty} \rd rr\re^{-z_nr}g_\textrm{HS}(r).
\end{equation}
Here, we will be using Percus-Yevick expression for $\widetilde{G}_\textrm{HS}(z_n)$, i.e.,
\begin{eqnarray}
\label{Gtran}
\widetilde{G}_{(\textrm{HS})}(z_n)=
\frac{[\lambda_2+z_n(\lambda_1+\lambda_2)]}{z_n^2\widetilde{D}_0^{(n)}} \, ,
\end{eqnarray}
where
\begin{eqnarray}
\label{lam12}
\lambda_1=-\frac{3}{2}\frac{\eta}{(1-\eta)^2} \,,\qquad \lambda_2=\frac{1+2\eta}{(1-\eta)^2}\,,
\end{eqnarray}
\begin{eqnarray}
\label{D0n}
\widetilde{D}_0^{(n)}=\left\{1-12\eta\left[\frac{\lambda_1+\lambda_2}{z_n^2}
\left(1-z_nd-\re^{-z_nd}\right)+
\frac{\lambda_2}{z_n^3}\bigg(1-z_nd+\frac{z_n^2}{2}-\re^{-z_nd}\bigg) \right] \right\}
\end{eqnarray}
and $\eta=\pi(\rho_{1}+\rho_{2})d^3/6$.

Differentiating the expression for Helmholtz free energy (\ref{FRE}) with respect to the density, we get the
following expression for the chemical potential:
\begin{equation}
\label{chempot}
\beta\mu_k^{(\textrm{HSY})}=\frac{\partial}{\partial\rho_k}\bigg(\frac{\beta A_\textrm{HSY}}{V}\bigg)=\beta\mu_k^{(\textrm{HS})}
+\beta\Delta\mu_k^{(\textrm{HSY})},
\end{equation}
where $\mu_k^\textrm{HSY}$ is the hard-sphere chemical potential and
\begin{eqnarray}
\label{chem000}
\beta\Delta\mu_k^{(\textrm{HSY})}=-\frac{2\pi\beta\epsilon_0}{z_n}\left[{\partial\widetilde{G}_\textrm{HS}(z_n)\over\partial\rho_k}
\sum_{i}\sum_{j}\rho_i\rho_jA_{ij}+2\widetilde{G}_\textrm{HS}(z_n)\sum_{i}\rho_{i}A_{ik}\right].
\end{eqnarray}
Pressure $P_\textrm{HSY}$ of the system can be calculated invoking the following general relation:
\begin{eqnarray}
\label{pressure}
\beta P_\textrm{HSY}=\beta \sum_{k}\rho_k\mu_k^{(\textrm{HSY})}-\frac{\beta A_\textrm{HSY}}{V}\,.
\end{eqnarray}

In the above expressions, $A_\textrm{HS}$ and $\mu_k^{(\textrm{HS})}$ are calculated using the corresponding Carnahan-Starling expressions \cite{CS}.

\subsection{Thermodynamic perturbation theory}

According to the TPT-CF for the associative part of the free energy $A_\textrm{ass}$, we have:
\begin{eqnarray}
\label{FRE2}
\frac{\beta A_\textrm{ass}}{V}=\sum_{k}\left[\rho_k
\ln \left(\frac{\sigma^{(0)}_{k}}{\rho_k}\right)+
\frac{1}{2}\sigma^{(m-1)}_{k}\frac{\sigma^{(1)}_{k}-\sigma^{(0)}_{k}}
{\sigma^{(0)}_{k}} \right],
\end{eqnarray}
where
\begin{eqnarray}
\label{sigl}
\sigma^{(l)}_{k}=\sigma^{(0)}_{k} \sum_{n=0}^{l}\frac{1}{n!}\left(\frac{\sigma^{(1)}_{k}-\sigma^{(0)}_{k}}
{\sigma^{(0)}_{k}}\right)^n \qquad  \text{for} \qquad  l=2,\ldots , m\,.
\end{eqnarray}
Here, $m$ is the maximum number of associative bonds per particle (the maximum number of the particles,
which can be bonded to a given particle simultaneously), $\sigma^{(l)}_k=\sum_l\rho^{(l)}_k$,
$\rho_k=\sum_l\rho_k^{(l)}$ and $\rho^{(l)}_k$ is the density of $l$-times bonded particles.
For the present two-component mixture, the density parameters $\sigma_k^{(0)}$ and $\sigma_k^{(1)}$
satisfy the following set of equations
\begin{equation}
\label{systeq}
\left\{\begin{array}{ll}
  \frac{\sigma_{1}^{(1)}-\sigma_{1}^{(0)}}{\sigma_{1}^{(0)}}=\left[\rho_2 -
  \frac{1}{m!}\frac{\left(\sigma_{2}^{(1)}-\sigma_{2}^{(0)}\right)^m}{
\left(\sigma_{2}^{(0)}\right)^{m-1}}\right]K,\\[2ex]
  \rho_{1}\left(\sigma_{1}^{(0)}\right)^{m-1}=\sum_{k=0}^{m}\frac{(\sigma_{1}^{(0)})^{k}}{(m-k)!}
\left(\sigma_{1}^{(1)}-\sigma_{1}^{(0)}\right)^{m-k} ,\\[2ex]
  \frac{\sigma_{2}^{(1)}-\sigma_{2}^{(0)}}{\sigma_{2}^{(0)}}=\left[\rho_1-
  \frac{1}{m!}\frac{\left(\sigma_{1}^{(1)}-\sigma_{1}^{(0)}\right)^m}{
\left(\sigma_{1}^{(0)}\right)^{m-1}}\right]K ,\\[2ex]
  \rho_{2}\left(\sigma_{2}^{(0)}\right)^{m-1}=\sum_{k=0}^{m}
\frac{\left(\sigma_{2}^{(0)}\right)^{k}}{(m-k)!}
\left(\sigma_{2}^{(1)}-\sigma_{2}^{(0)}\right)^{m-k},              \end{array}
                   \right.
\end{equation}
where
\begin{equation}
\label{K}
K=4\pi\int y^{(00)}_{12}(r)e^{(\textrm{HSY})}(r)f_\textrm{ass}(r)r^2\rd r=4\pi BL^2y^{(00)}_{12}(L),
\end{equation}
 $y^{(00)}_{12}(r)$ represent the cavity distribution function between two Yukawa hard
spheres of species 1 and 2 infinitely diluted in the original associating fluid in question.
Usually, this function is approximated by the hard-sphere Yukawa cavity correlation function
$y_{12}^{(\textrm{HSY})}(r,\eta)$
calculated for the packing fraction $\eta$. This appears to be a good approximation for the models with
bonding length $L\approx d$, since in this case $y^{(00)}_{12}(r)$ only weakly depends on the
degree of the system association. However, for $L<d$, the actual (effective) packing fraction $\eta_\textrm{eff}$ and
thus $y^{(00)}_{12}(r)$ are strongly dependent on the system degree of association, and the usual approximation
becomes inadequate. In the present study we propose to approximate $y^{(00)}_{12}(r)$ by the hard-sphere
Yukawa cavity correlation function $y_{12}^{(\textrm{HSY})}(r,\eta_\textrm{eff})$ calculated for the effective packing fraction $\eta_\textrm{eff}$,
i.e.,
\begin{eqnarray}
\label{etaeff}
\eta_\textrm{eff}=\frac{\pi d^3}{6}\sum_k\rho_kX_k^{(0)}
+\sum_{n=1}^{m}
\left(\frac{\pi d^3}{6}-n V_\textrm{exc}\right)
\sum_k
\frac{\rho_k X^{(0)}_{k}}{n!}\left(\frac{X^{(1)}_{k}}{X^{(0)}_{k}}\right)^{n},
\end{eqnarray}
where the excluded volume $V_\textrm{exc}$ is:
\begin{eqnarray}
\label{exvol}
V_\textrm{exc}=\frac{\pi}{24}(d-L)^{2}(2d+L).
\end{eqnarray}
$X_k^{(0)}=\rho_k^{(0)}/\rho_k$ and $X_k^{(1)}=\rho_k^{(1)}/\rho_k$. According to this expression, $\eta_\textrm{eff}$
[and thus $y_{12}^{(\textrm{HSY})}(r)$] depends on the degree of association of the system represented by the fractions
of free $X_k^{(0)}$ and singly bonded $X_k^{(1)}$ particles.

In the present study, the solution of this equation is obtained via numerical iteration method. On each iteration
step, the new estimate for the fractions $X^{(l)}_{k,\textrm{new}}$ $(l=0,1)$ is calculated by solving the following set of
equations:
\begin{equation}
\label{systeq1}
\left\{\begin{array}{ll}
\frac{X_{1,\textrm{new}}^{(1)}}{X_{1,\textrm{new}}^{(0)}}=\left[1-
\frac{1}{m!}\frac{\left(X_{2,\textrm{new}}^{(1)}\right)^m}{\left(X_{2,\textrm{new}}^{(0)}\right)^{m-1}}\right]\rho_2 K\left[\eta_\textrm{eff}(X_\textrm{old})\right],\\[3ex]
\left(X_{1,\textrm{new}}^{(0)}\right)^{m-1}=\sum_{n=0}^{m}\frac{\left(X_{1,\textrm{new}}^{(0)}\right)^{n}}{(m-n)!}
\left(X_{1,\textrm{new}}^{(1)}\right)^{m-n},\\[2ex]
\frac{X_{2,\textrm{new}}^{(1)}}{X_{2,\textrm{new}}^{(0)}}=\left[1-
\frac{1}{m!}\frac{\left(X_{1,\textrm{new}}^{(1)}\right)^m}{\left(X_{1,\textrm{new}}^{(0)}\right)^{m-1}}\right]\rho_1 K\left[\eta_\textrm{eff}(X_\textrm{old})\right],\\[3ex]
\left(X_{2,\textrm{new}}^{(0)}\right)^{m-1}=\sum_{n=0}^{m}\frac{\left(X_{2,\textrm{new}}^{(0)}\right)^{n}}{(m-n)!}
\left(X_{2,\textrm{new}}^{(1)}\right)^{m-n},
\end{array}
                   \right.
\end{equation}
which is obtained using a set of equations (\ref{systeq}). Here, $X_\textrm{old}$ is the value of  $X$ calculated during the previous iteration step.
Our iteration loop consists of two steps. In the first step, the current value of $\eta_\textrm{eff}$ is used
to calculate the new values of $X_k^{(l)}$ using the set of equation (\ref{systeq1}). On the second step, we insert
these values of $X_k^{(l)}$ into the right-hand side of the relation (\ref{etaeff})
to get a new estimate for $\eta_\textrm{eff}$. This iteration loop is repeated until the following condition
\begin{equation}
{\frac{|\eta_{\textrm{eff},\textrm{new}}-\eta_\textrm{eff,old}|}{ \eta_{\textrm{eff},\textrm{new}}+\eta_{\textrm{eff,old}}}}\leqslant 10^{-8}
\end{equation}
is satisfied. For the initial guess we have used the value of $\eta_\textrm{eff}=\eta$.

\subsection{The cavity correlation function for Yukawa hard sphere fluid}

The cavity correlation function $y_{12}^{(\textrm{HSY})}(r)$, which is needed to solve the set of equations for the
fractions $X_k^{(l)}$ (\ref{systeq1}) is calculated using the reference hypernetted chain type of
approximation
\begin{equation}
\label{gr}
y_{ij}^{(\textrm{HSY})}(r)=y_{ij}^{(\textrm{HS})}(r)\exp{\left[\delta h_{ij}^{(\textrm{HSY})}(r)-\delta c_{ij}^{(\textrm{HSY})}(r)\right]},\;\;\;\;\;\;
\end{equation}
where $y_{ij}^{(\textrm{HS})}(r)$ is the hard-sphere cavity correlation function,
$\delta h_{ij}^{(\textrm{HSY})}(r)=h_{ij}^{(\textrm{HSY})}(r)-h_{ij}^{(\textrm{HS})}(r)$ and
$\delta c_{ij}^{(\textrm{HSY})}(r)=c_{ij}^{(\textrm{HSY})}(r)-c_{ij}^{(\textrm{HS})}(r)$. Here, the upper indices (HS) and (HSY)
denote the hard-sphere and hard-sphere Yukawa quantities, respectively, and $h$ and $c$ denote total and
direct correlation functions, respectively. In the hard-core region $\delta h_{ij}^{(\textrm{HSY})}(r)=0$ and
for $\delta c_{ij}^{(\textrm{HSY})}(r)$, we have used the expression obtained in the framework of the first-order mean
spherical approximation \cite{MSA1}. The hard-sphere cavity correlation function $y_{ij}^{(\textrm{HS})}(r)$ was
calculated using Henderson-Grundke approximation \cite{GH}. Closed form analytical expressions for
$\delta c_{ij}^{(\textrm{HSY})}(r)$ and $y_{ij}^{(\textrm{HS})}(r)$ are presented in the appendix.

\subsection{Calculation of the phase diagram}

Our calculation of the phase diagram follows closely the scheme, proposed in \cite{kahl1}.
It is based on the solution of the set of equations that follow from the conditions of phase equilibrium, i.e., equal chemical potentials and pressures of the coexisting phases at a given temperature. The coexisting phases are characterized by $(\rho,x)$ and $(\rho',x')$. From the Gibbs' phase rule, we expect up to four phases to be in equilibrium, i.e., the vapour (V), the mixed fluid (MF), and two (symmetric) phases of the demixed fluid (DF).

The V-MF transition is obtained by solving the set of equations:
\begin{equation}
\label{gibb1}
\mu_i(\rho,T,x=1/2)\equiv\mu(\rho,T,x=1/2)=\mu(\rho',T,x=1/2),
\end{equation}
\begin{equation}
\label{gibb2}
P(\rho,T,x=1/2)=P(\rho',T,x=1/2).
\end{equation}
The V-MF and MF-DF transitions are obtained in two steps: first we determine the phase diagram of the demixing transitions , i.e., looking  at a given temperature $T$ for two coexisting states with the same fluid density but different composition by fixing $\rho=\rho'$  and by determining concentrations $x$ and $x'=1-x$ of the coexisting phases. The equilibrium condition for the pressure is automatically fulfilled, while the equilibrium condition for the chemical potentials takes place at given $T$ and $\rho$
\begin{equation}
\label{gibb3}
\mu(\rho,T,x)=\mu(\rho,T,x),
\end{equation}
which defines the line $x(\rho)$ of the second order transition.

In the second step, the solution of the two equations
\begin{align}
\label{gibb4}
\mu[\rho,T,x=1/2]&=\mu[\rho',T,x(\rho')],\\
\label{gibb5}
P[\rho,T,x=1/2]&=P[\rho',T,x(\rho')]
\end{align}
gives the density $\rho$ of the V or MF and the density of the DF with concentrations $x(\rho')$ and $1-x(\rho')$, in equilibrium.

\section{Results and discussion}

\begin{figure}[!t]
\centerline{
\includegraphics[clip=true,width=0.5\textwidth]{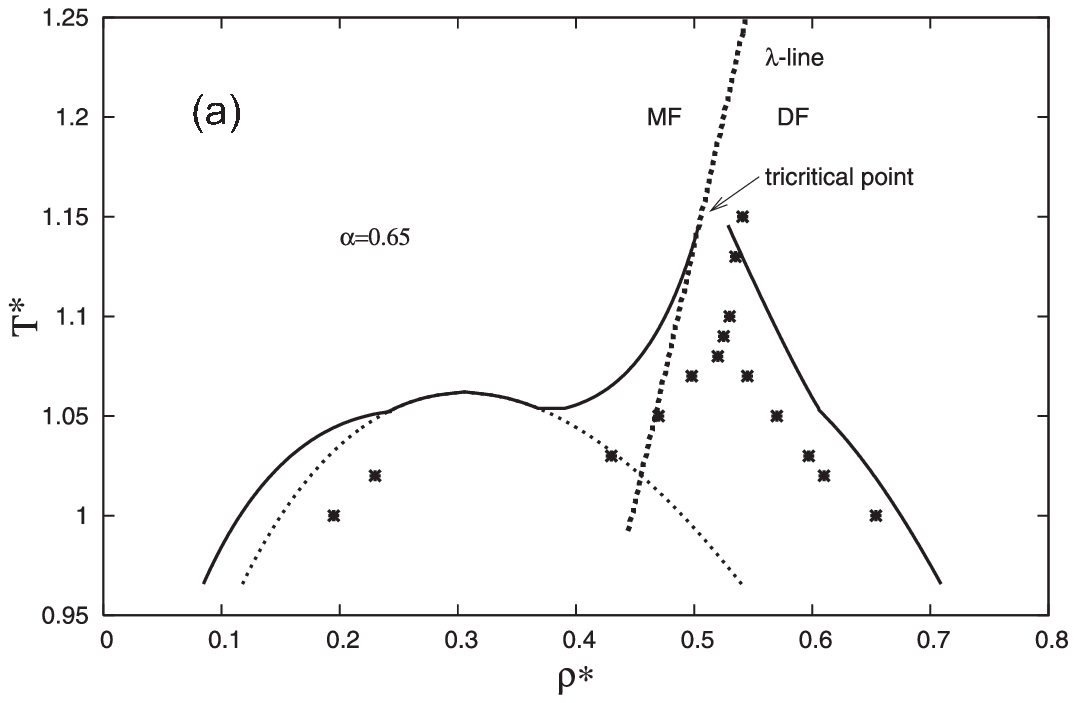}
\includegraphics[clip=true,width=0.5\textwidth]{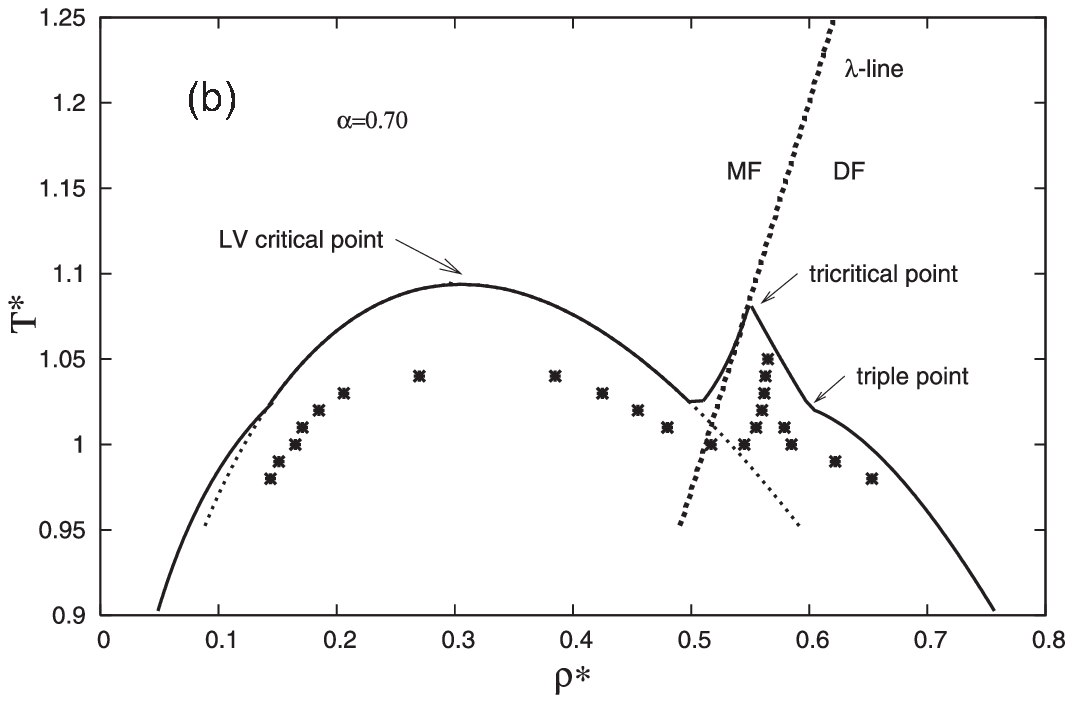}
}
\centerline{
\includegraphics[clip=true,width=0.5\textwidth]{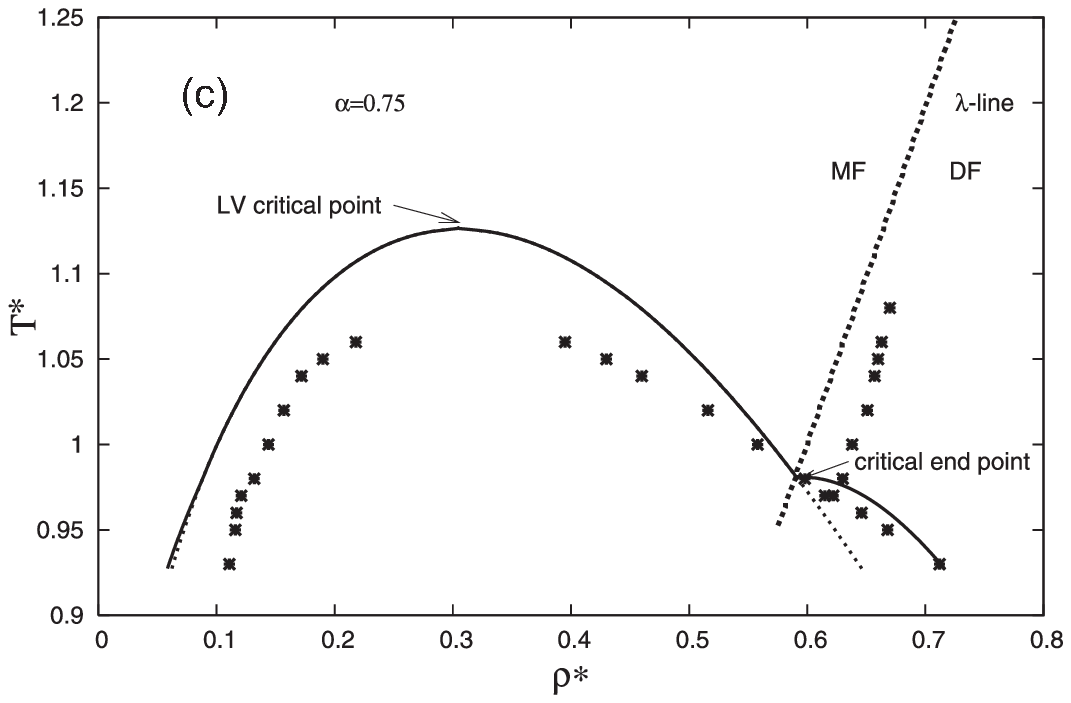}
}
\caption{Phase diagram of the symmetric binary Yukawa hard-sphere associating mixture in
$\rho^*$ vs $T^*$ coordinate frame for $\epsilon^*=0$ and $\alpha=0.65$ (panel a),
$\alpha=0.7$ (panel b) and $\alpha=0.75$ (panel c). Lines represent the results
of the present theory and symbols depict MC computer simulation results \cite{kahl3}.
Here, dashed lines denote the $\lambda$-lines and dotted lines show the
metastable LV binodals.}
\label{X1}
\end{figure}

\begin{figure}[!b]
\centerline{
\includegraphics[clip=true,width=0.5\textwidth]{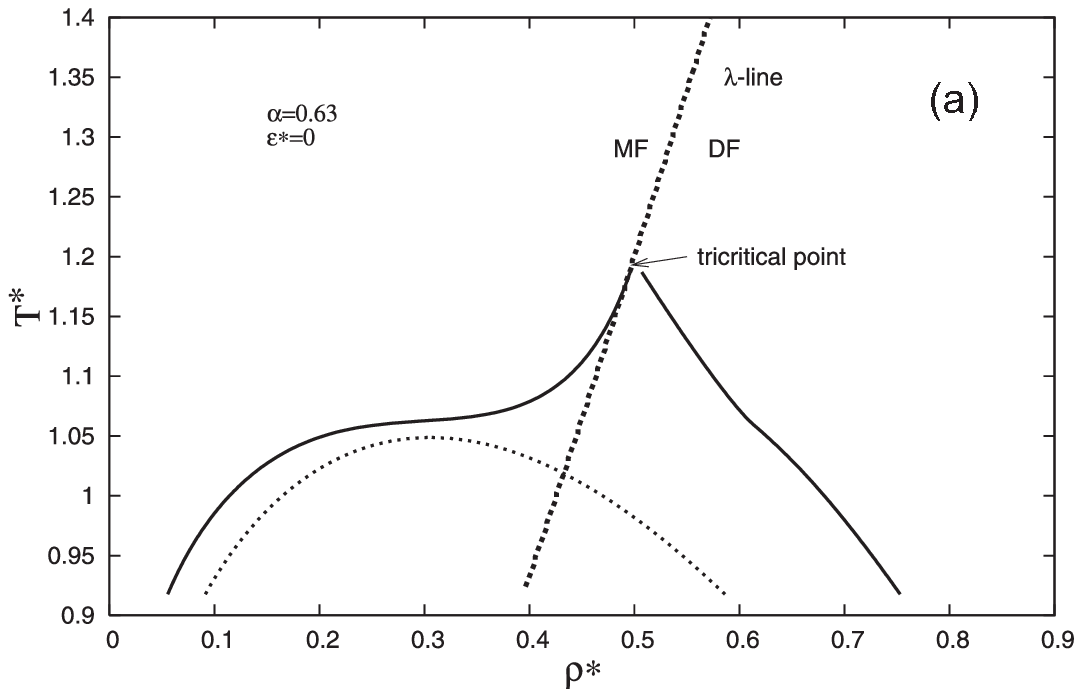}
\includegraphics[clip=true,width=0.5\textwidth]{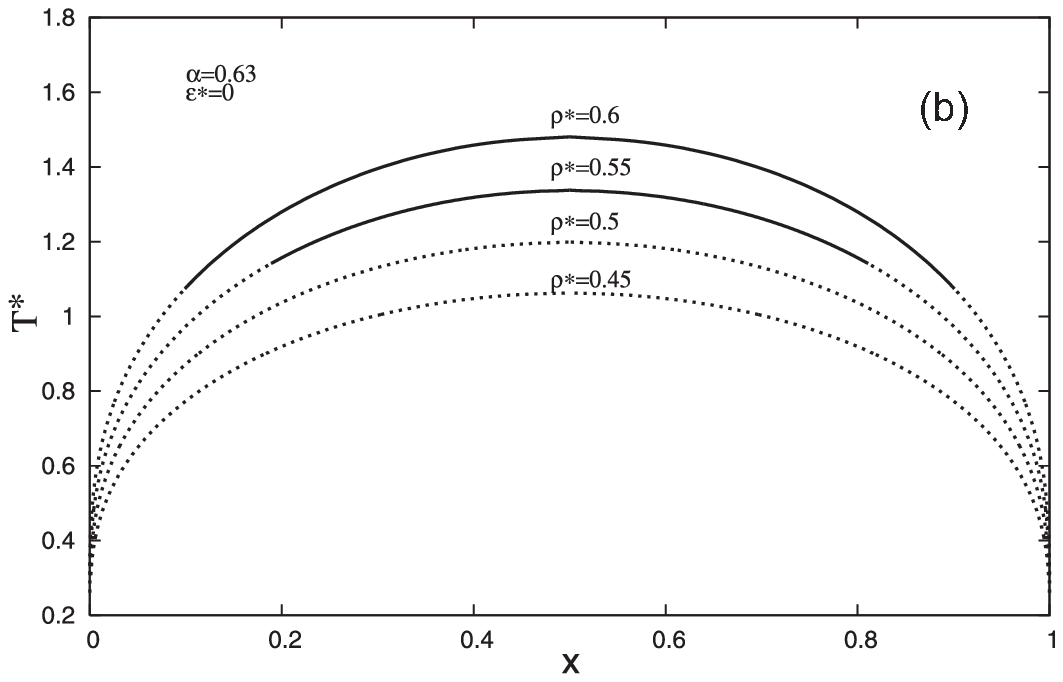}
}
\caption{Phase diagram of the symmetric binary Yukawa hard-sphere associating mixture in
$\rho^*$ vs $T^*$ coordinate frame (panel a) and in $x$ vs $T^*$ frame for different values of the
density $\rho^*$ (panel b)
at $\epsilon^*=0$ and $\alpha=0.63$. Dashed lines denote the $\lambda$-line and solid and dotted lines represent stable and unstable portions of the coexisting densities (panel a) and coexisting mole (number) fractions (panel b), respectively.}
\label{X2}
\end{figure}

In this section we present our numerical results for the phase behavior of the model in question. All the
calculations are carried out at Yukawa screening parameter $z_{n}d=1.8$ and square-well width $\omega=0.0000404981$.

\begin{figure}[!t]
\centerline{
\includegraphics[clip=true,width=0.5\textwidth]{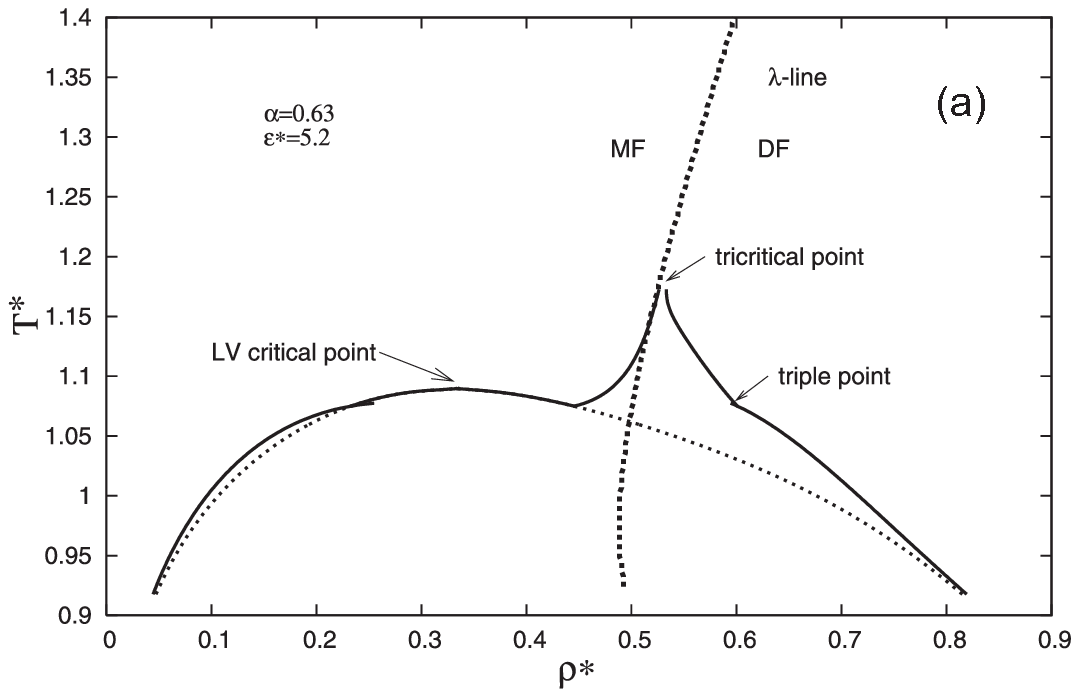}
\includegraphics[clip=true,width=0.5\textwidth]{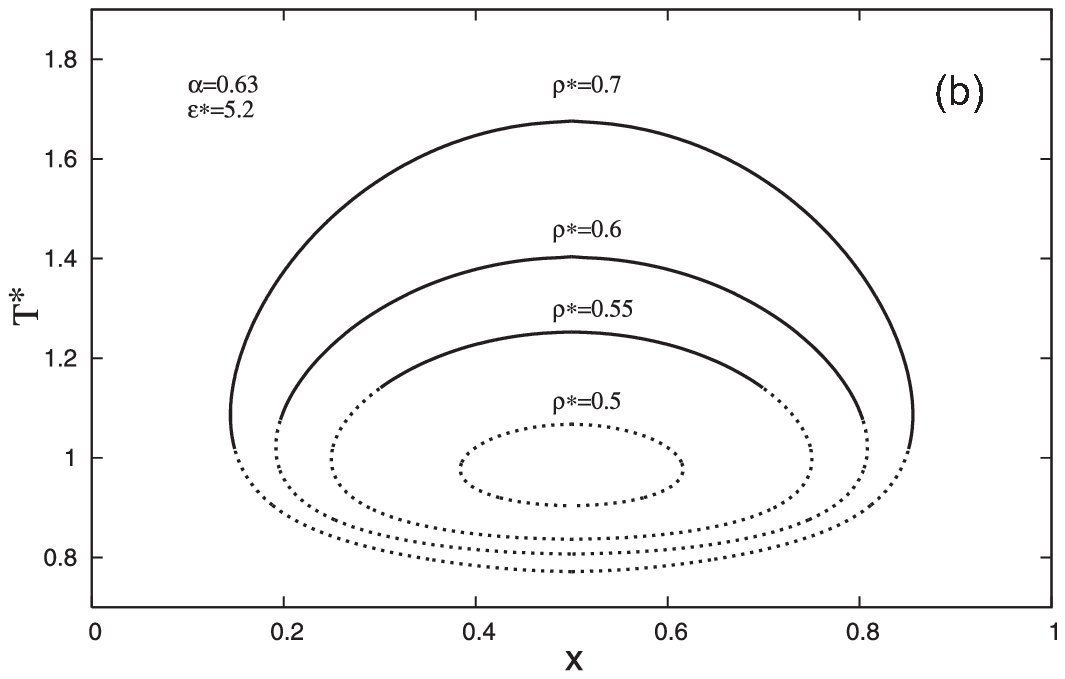}
}
\caption{The same as in figure 2 at $\epsilon^*=5.2$ and $\alpha=0.63$.}
\label{X3}
\end{figure}

According to the previous studies \cite{kal_dim}, predictions of our theory for thermodynamical properties of the
model with $A_{ij}=0$ are in a good agreement with computer simulation predictions.
To test the accuracy of the theory for the model with $\epsilon_\textrm{ass}^*=0$,
we compare theoretical and computer simulation predictions for its phase behavior.
In figure~\ref{X1} we show the phase diagram of the system at $\epsilon_\textrm{ass}^*=0$ and three values of $\alpha$,
i.e., $\alpha=0.65,\;0.7,\;0.75$. These are the system parameters for which the three types of the phase
diagram were identified \cite{wilding,kahl1}, depending on the position of intersection point of the
$\lambda$-line, which represent the second-order demixing transition, with the binodals of the
liquid-vapour
 (LV) phase transition. In addition, for comparison in the same figure, we present the corresponding computer
simulations results \cite{kahl2}. Overall there is a reasonably good qualitative agreement between
theoretical and computer simulation predictions. Predictions of the theory in the region of the LV
critical point are about $6\%$ higher than those of the computer simulation. As a result, while the
types I and II of the phase diagrams (according to the nomenclature of references \cite{wilding,kahl1}) are
theoretically reproduced for the set of the potential model parameters used to simulate the
type III of the diagram, theoretical
calculations still show the type II of the diagram with a small portion of stable binodals in the
vicinity of the LV critical point. However, it is quite obvious that a small decrease
in $\alpha$ will cause the theoretical phase diagram to change its type from type II to type III.
This can be seen in figure~\ref{X2} (panel a), where our results for $\alpha=0.63$ are shown.
In the phase diagram of the type I the LV, coexistence is unstable with respect to the
three-phase coexistence between mixed fluid (MF) and demixed fluid (DF) and the $\lambda$-line ends
at the tricritical point. In the type II of the diagram, $\lambda$-line ends also at the tricritical point,
however, there is a portion of the LV phase diagram beeing stable in the range of the temperatures
between the critical temperature and the temperature of the triple point, where one can observe LV
coexistence at lower densities. Between tricritical temperature and temperature of the triple point,
the MF-DF three-phase coexistence can be seen. At the triple point, there is a four-phase vapour, MF and DF coexistence.
In the case of the type II diagram, the $\lambda$-line intersects the liquid binodal at the temperature
slightly below the critical. In the type III of the diagram, the $\lambda$-line intersects the liquid
binodal at the temperature well below the critical temperature. Here, we can see the critical end point
below which there is a three phase V-DF coexistence and above which (up to the LV critical temperature), there is a LV coexistence. In the type IV of the diagram, the $\lambda$-line intersects LV binodals at the
densities that are lower than the LV critical densities \cite{kahl3}. This occures at $\alpha=0$. We
 have also detected this type of the diagram, using the current approach; however, the results are not shown here.

\begin{figure}[!b]
\centerline{
\includegraphics[clip=true,width=0.5\textwidth]{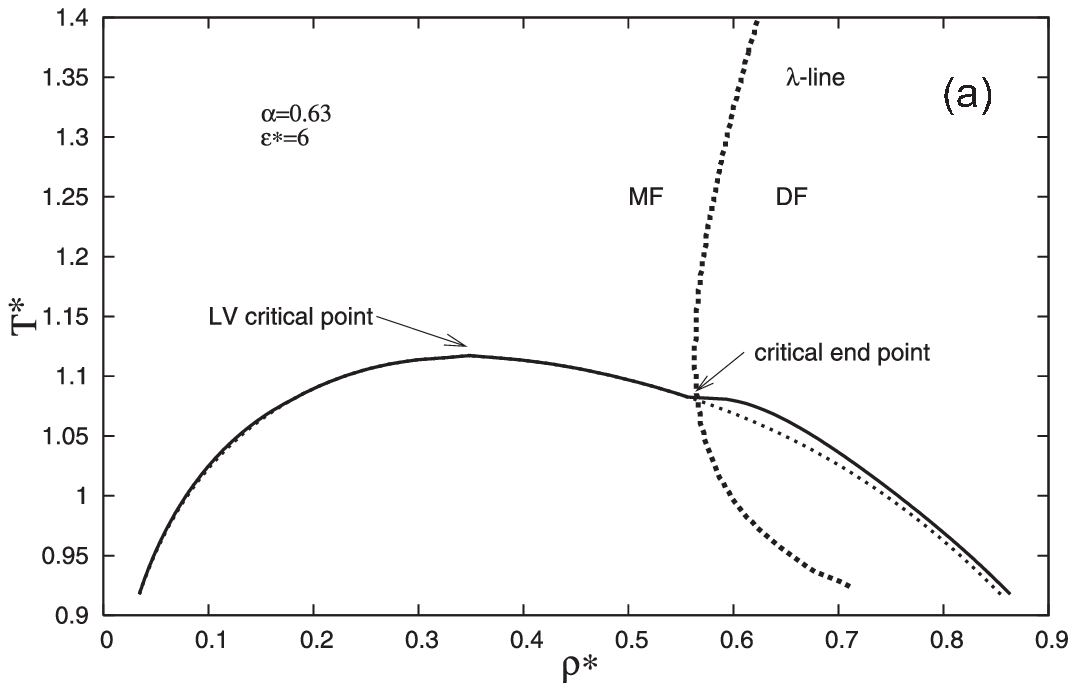}
\includegraphics[clip=true,width=0.5\textwidth]{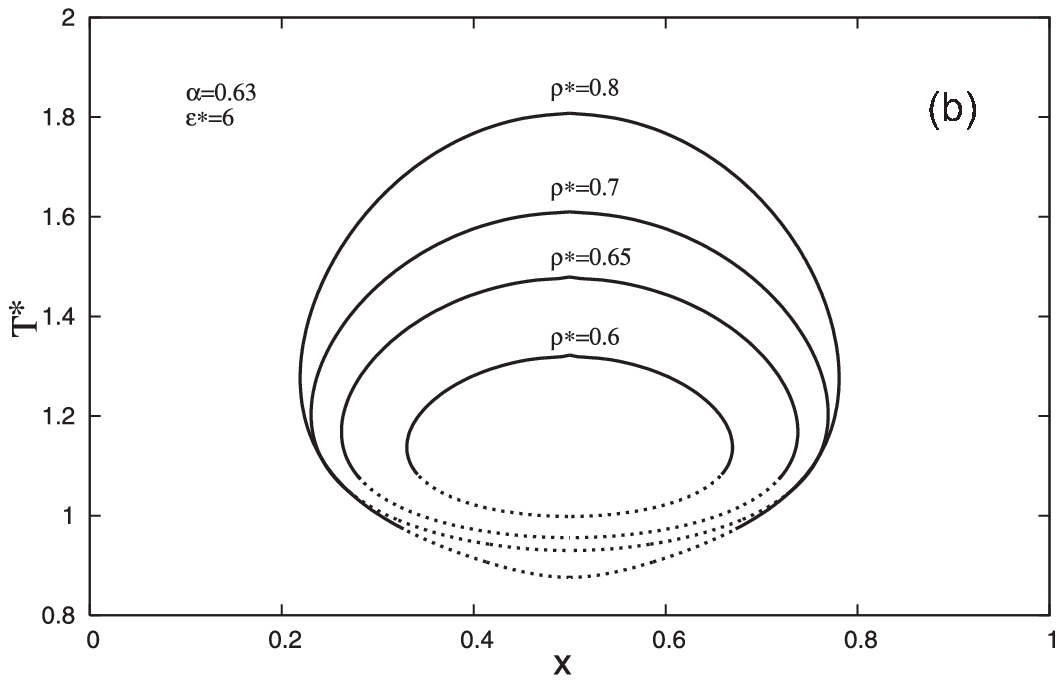}
}
\caption{The same as in figure 2 at $\epsilon^*=6$ and $\alpha=0.63$.}
\label{X4}
\end{figure}

\begin{figure}[!t]
\centerline{
\includegraphics[clip=true,width=0.5\textwidth]{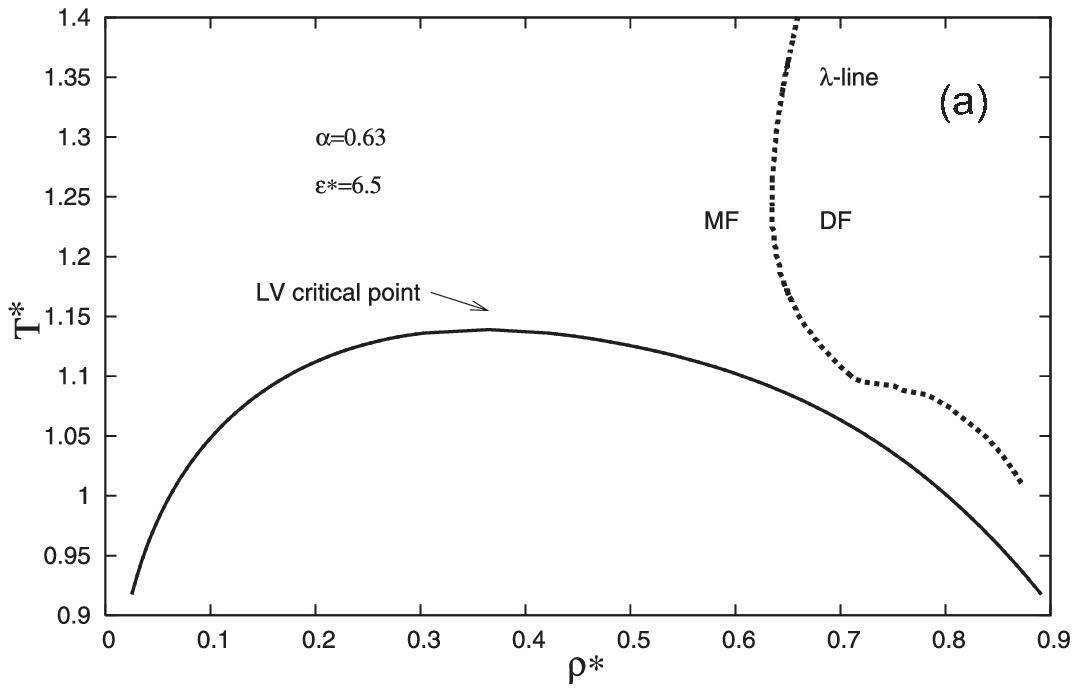}
\includegraphics[clip=true,width=0.5\textwidth]{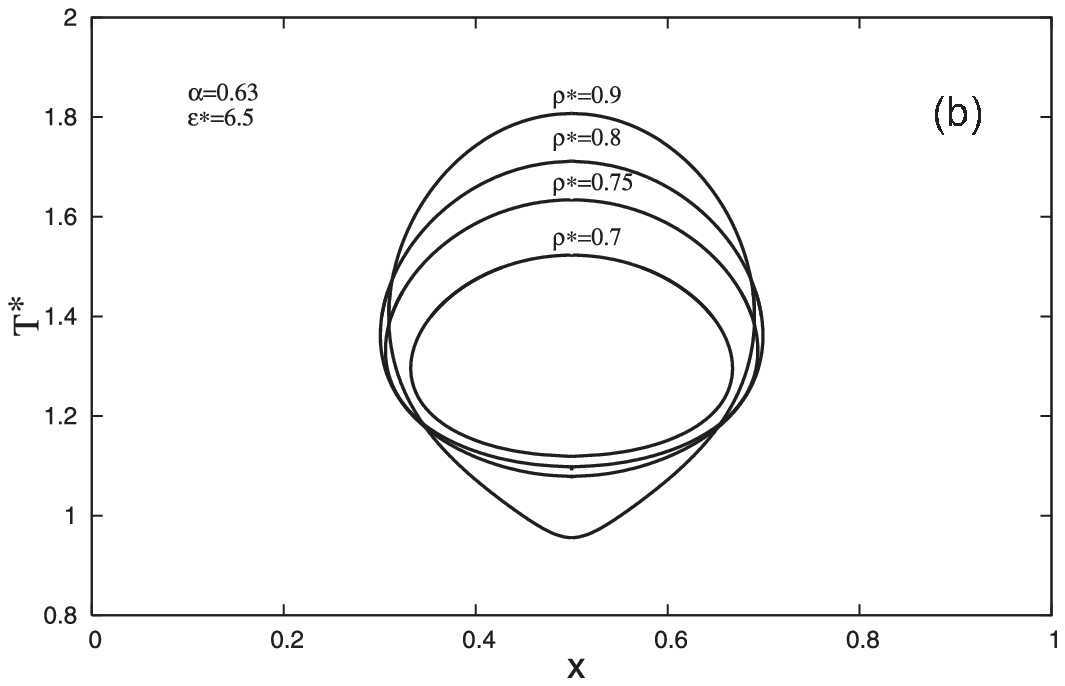}
}
\caption{The same as in figure 2 at $\epsilon^*=6.5$ and $\alpha=0.63$.}
\label{X5}
\end{figure}

Next, we proceed to the discussion of the phase diagrams for the nonzero value of the strength of
associating interaction $\epsilon^*=5.2,\;6.0,\;6.5$ at $\alpha=0.63$ (figures~\ref{X3}--\ref{X5}).
Unfortunately, computer simulation results for the phase behavior of the model at hand are not available.
However, taking into account a reasonable performance of the theory in the two limiting cases discussed
above ($A_{ij}=0$ and $\epsilon_0=0$), we expect that the accuracy of the theory for the full version
of the model will be satisfactory as well. In figures~\ref{X3}--\ref{X5},
we depict the phase diagram in $\rho^*$ vs $T^*$ (panel a) and  $x$ vs $T^*$ at different values of the density (panel b) frames. For $\epsilon=0$ and $\alpha=0.63$, temperature-concentration slices of the phase diagram
at different densities are also shown (panel b in figure~\ref{X2}). In the latter case, only the upper
portions of the corresponding coexistence curves for $\rho^*=0.55,\;0.6$ are stable. The lower portions of these curves and the demixing curves for $\rho^*=0.5,\;0.45$ are unstable with respect to the three-phase
MF-DF coexistance.
With the
temperature decrease, the difference in the compositions of the coexisting liquids increases. With the
increase of the strength of association $\epsilon^*$, the topology of the phase behavior in $T^*$ vs
$\rho^*$ coordinate frame changes from the type III (figure~\ref{X2}) to type II at $\epsilon^*=5.2$
(figure~\ref{X3}) and next to type I at $\epsilon^*=6.0$ (figure~\ref{X4}). At the same time, one can observe the
appearance of the closed loop liquid-liquid immiscibility curves with the upper stable and lower unstable
critical solution points (figures~\ref{X3} and \ref{X4}). The stable portion of the curves increases with an increasing strength
of associative interaction. Finally, for $\epsilon^*=6.5$, the closed-loop coexistence curves
for demixing coexistence becomes stable (figure~\ref{X5}, panel b). This corresponds to the situation
when there is no intersection between LV binodals a $\lambda$-line (figure~\ref{X5},  panel a).
Thus, for these values of the potential model parameters, in addition to already identified four types
of the phase diagram, we have identified one more, which we call the type V of the two-component
mixture phase diagram topology. In a future we are planning to extend and apply our approach
to study the effects of the external field \cite{kahl5} and porous media \cite{hol1,hol2} on the phase behaviour of the current model.

\section{Conclusions}

In this paper we have used the TPT-CF approach to study the phase behavior
of a symmetric two-component Yukawa mixture of associating particles with spherically symmetric interaction.
Our theoretical predictions for the phase diagram of the version of the model without association
appear to be in reasonably good
qualitative agreement with the predictions of the corresponding Monte-Carlo computer simulation method \cite{kahl2}. For the model with nonzero associating potential, we were able to identify, in addition to
the already known three types of the phase diagram topologies \cite{wilding,kahl1}, the type V of the phase
diagram. This type is characterized by the absence of intersection of the $\lambda$-line, which
represents a demixing coexistence, with LV binodals. As a result, the stable closed-loop liqui-liquid
immiscibility curves with upper and lower critical solution temperatures can be observed for the larger
values of the temperature and density. Thus, closed-loop liquid-liquid immiscibility, which was observed earlier for the binary systems with highly directional attractive forces \cite{jackson}, can be also seen  for
the binary fluids with spherically symmetric interaction.

\appendix


\section{Grundke-Henderson approximation}

\renewcommand{\theequation}{A.\arabic{equation}}

To calculate the hard-sphere cavity correlation function $y_{ij}^{(\textrm{HS})}(r)$ we use Grundke-Henderson approximation \cite{GH}. For $r<d$, we have:
\begin{equation}
\label{lnyr}
\ln y_{ij}^{(\textrm{HS})}(r)=\sum_{n=0}^{3}a_nr^n,
\end{equation}
where $a_0$ and $a_1$ are determined from (\ref{lny0}) and (\ref{dlnyr0}), respectively, and $a_2$ and $a_3$ are determined by requiring that $y_{ij}^{(\textrm{HS})}(r)$ and ${\partial y_{ij}^{(\textrm{HS})}(r)}/{\partial r}$ should be continuous at $r=d$ [(\ref{ysig})and (\ref{dysig})].
\begin{equation}
\label{lny0}
\ln y_{ij}^{(\textrm{HS})}(0)=\frac {8\eta-9\eta^2+3\eta^3}{(1-\eta)^3}\,,
\end{equation}
\begin{equation}
\label{dlnyr0}
\bigg(\frac{\partial \ln y_{ij}^{(\textrm{HS})}(r)}{\partial r}\bigg)_{(r=0)}=-3\eta\frac{2-\eta}{(1-\eta)^3}\,,
\end{equation}
\begin{equation}
\label{ysig}
y_{ij}^{(\textrm{HS})}(d)=\frac {4-2\eta}{4(1-\eta)^3}\,,
\end{equation}
\begin{equation}
\label{dysig}
\bigg(\frac{\partial y_{ij}^{(\textrm{HS})}(r)}{\partial r}\bigg)_{(r=d)}=\frac{5\eta^2-9/2\eta}{(1-\eta)^3}\,.
\end{equation}
Thus, we obtain $a_0$, $a_1$, $a_2$, $a_3$ by combining equations (\ref{lnyr}), (\ref{lny0}), (\ref{dlnyr0}), (\ref{ysig}) and (\ref{dysig}):
\begin{eqnarray}
\label{a0a1a2a3}
&&a_0=\frac {8\eta-9\eta^2+3\eta^3}{(1-\eta)^3}\,, \qquad\qquad
a_1=-3\eta\frac{2-\eta}{(1-\eta)^3}\,,\nonumber\\
&&a_2=\frac{1}{d^2}\left\{-\left(3a_0+2a_1d\right)+3\ln\left[\frac{4-2\eta}{4(1-\eta)^3}\right]+
\frac{9\eta+10\eta^2}{2-\eta}\right\}\,,\nonumber\\
&&a_3=\frac{1}{d^3}\left\{2a_0+a_1d-2\ln\left[\frac{4-2\eta}{4(1-\eta)^3}\right]-
\frac{9\eta+10\eta^2}{2-\eta}\right\}\,.
\end{eqnarray}

\section{First-order mean spherical approximation}
\renewcommand{\theequation}{B.\arabic{equation}}

Using first-order mean spherical approximation \cite{MSA1}, for $r<1$, we have:
\begin{eqnarray}
\label{c1r}
\delta c_{ij}^{(\textrm{HSY})}(r)&=&\beta\frac{\epsilon_0}{r}\frac{A_{ij}}{z_n}\re^{-z_n(r-d)}
-\frac{\beta\epsilon_0}{r}\frac{A_{ij}}{z_n}
\frac{1}{(1-\eta)^4 z_n^6Q_0^2(z_n)} \nonumber\\
&&\times \Big\{S^2(z_n)\re^{-z_n(r-d)}+144\eta^2L^2(z_n)\re^{z_n(r-d)}+24\eta S(z_n)L(z_n)
\nonumber\\&&
-12\eta^2\left[(1+2\eta)^2z_n^4+(1-\eta)(1+2\eta)z_n^5\right]r^4
\nonumber\\[1ex]&&
+12\eta\left[S(z_n)L(z_n)z_n^2-(1-\eta)^2(1+\eta/2)z_n^6\right]r^2
\nonumber\\&&
-24\eta\left[(1+2\eta)^2z_n^4+(1-\eta)(1+2\eta)z_n^5\right]r\Big\},
\end{eqnarray}
where
\begin{eqnarray}
\label{Q0zfmsa}
Q_0(z_n)=\frac{S(z_n)+12\eta L(z_n)\re^{-z_n}}{(1-\eta)^2z_n^3}\,,
\end{eqnarray}
\begin{eqnarray}
\label{Szfmsa}
S(z_n)=(1-\eta)^2z_n^3+6\eta(1-\eta)z_n^2+18\eta^2z_n-12\eta(1+2\eta)\,,
\end{eqnarray}
\begin{eqnarray}
\label{Lzfmsa}
L(z_{n})=\bigg(1+\frac{\eta}{2}\bigg)z_{n}+1+2\eta\,.
\end{eqnarray}



\ukrainianpart

\title{Фазовий перехід ``рідина-рідина'' із замкнутою областю незмішування у суміші сферично-симетричних частинок}
%
%
\author{Ю.В. Калюжний, Т.В. Гвоздь}
\address{Інститут фізики конденсованих систем НАН України, вул. І.~Свєнціцького, 1, 79011  Львів, Україна}

\makeukrtitle

\begin{abstract}
\tolerance=3000%
В рамках термодинамічної теорії збурень для асоціативного потенціалу типу центральних сил проведено дослідження фазової поведінки симетричної бінарної суміші асоціативних частинок з сферично-симетричною взаємодією. Модель представлено бінарною сумішшю юкавівських твердих сфер з додатковою сферично-симетричною асоціативною взаємодією типу прямокутної ями, яка розміщена усередині області твердої сфери  і діє тільки між різними сортами. Враховуючи зміну упаковки системи внаслідок асоціації, запропоновано узагальнення термодинамічної теорії збурень для асоціативного потенціалу типу центральних сил. На додаток до чотирьох вже відомих типів фазових діаграм для бінарних сумішей, нам вдалося визначити п'ятий тип, який характеризується відсутністю перетину лямбда-лінії з бінодалями ``рідина-газ'' і появою незмішування ``рідина-рідина'' у вигляді замкненої петлі з верхньою і нижньою критичними температурами змішування.

\keywords термодинамічна теорія збурень, співіснування ``рідина-газ'', розшаровування, бінарна суміш, асоціативні рідини

\end{abstract}

\end{document}